\documentstyle[11pt]{article}

\newcommand{\SI}{\Sigma}
\newcommand{\si}{\sigma}

\newcommand{\xn}{x_{n}}

\newcommand{\xm}{x_{m}}

\newcommand{\e}{e^{i k_{0}Y}}

\newcommand{\kim}{ k_{1}^{\mu}}                                      
\newcommand{\kom}{ k_{0}^{\mu}}                                      
\newcommand{\ki}{ k_{1}}
\newcommand{\yn}{ Y_{n}}                                             
 
\newcommand{\kn}{ k_{n}}
\newcommand{\kb}{ \bar{k}}

\newcommand{\km}{ k_{m}}

\newcommand{\kt}{ k_{2}}                                             
\newcommand{\ko}{ k_{0}}                                             
                                             
\newcommand{\yim}{ Y_{1}^{\mu}}                                      
\newcommand{\yin}{ Y_{1}^{\nu}}                                      
\newcommand{\kin}{ k_{1}^{\nu}}                                      
\newcommand{\kon}{ k_{0}^{\nu}}                                      
\newcommand{\ktm}{ k_{2}^{\mu}} 
\newcommand{\ktn}{ k_{2}^{\nu}}                                      
\newcommand{\ytm}{ Y_{2}^{\mu}}                                      
\newcommand{\ytn}{ Y_{2}^{\nu}}

\newcommand{\dsi}{\frac{\partial}{\partial x_{1}}}

\newcommand{\dsib}{\frac{\partial }{\partial \xb _{1}}}

\newcommand{\dst}{\frac{\partial }{\partial x_{2}}}             
\newcommand{\dstb}{\frac{\partial }{\partial \xb _{2}}}        
            
\newcommand{\dds}{\frac{\delta}{\delta \sigma}}

\newcommand{\dsb}{\frac{\partial ^{2}  }{\p x_1 \p \xb _1}}
\newcommand{\dsnm}{\frac{\partial ^{2}}{\partial x_{n}\partial x_{m}}}
\newcommand{\dsnmb}{\frac{\partial ^{2}}{\partial x_{n}\partial \xb _{m}}}
\newcommand{\dsii}{\frac{\partial ^{2}}
{\partial x_{1}^{2}}}
\newcommand{\p}{\partial}                                           
\newcommand{\pb}{\bar \partial}                                           
\newcommand{\pp}{\partial ^{2}}

\newcommand{\li}{ \lambda_{1}}

\newcommand{\al}{\alpha }

\newcommand{\alb}{\mbox{$\bar{\alpha }$}}
\newcommand{\xb}{\mbox{$\bar{x}$}}

\newcommand{\zb}{\mbox{$\bar{z}$}}    
\newcommand{\tb}{\mbox{$\bar{t}$}}   
\newcommand{\ai}{\mbox{$\alpha _{1}$}}

\newcommand{\aib}{\mbox{$\bar{\alpha _{1}}$}} 
\newcommand{\at}{\mbox{$\alpha _{2}$}} 
\newcommand{\atb}{\mbox{$\bar{\alpha _{2}}$}}

\newcommand{\kimb}{\mbox {$ \bar{k_{1}^{\mu}}$}} 
\newcommand{\kinb}{\mbox {$ \bar{k_{1}^{\nu}}$}}

\newcommand{\kib}{\mbox {$ \bar{k_{1}}$}} 
\newcommand{\ktb}{\mbox {$ \bar{k_{2}}$}} 
\newcommand{\qt}{\mbox {$ q_{2}$}} 
\newcommand{\qi}{\mbox {$ q_{1}$}}
\newcommand{\qtb}{\mbox {$ \bar{q_{2}}$}}
\newcommand{\qib}{\mbox {$ \bar{q_{1}}$}} 
\newcommand{\qo}{\mbox {$ q_{0}$}}

\newcommand{\la}{\mbox{$ \lambda $}} 
\newcommand{\be}{\begin{equation}}
\newcommand{\br}{\begin{eqnarray}}
\newcommand{\ee}{\end{equation}} 
\newcommand{\er}{\end{eqnarray}}

\newcommand{\dsiib}{\frac{\partial ^{2}} 
{\partial \xb _{1}^{2}}}          
\newcommand{\bj}{\mbox {$ \bar{\partial ^{2}}$}} 
\newcommand{\bjj}{\mbox {$ \bar{\partial ^{3}}$}}

\newcommand{\ppp}{\mbox {$ \partial ^{3}$}}

\begin{document}
\renewcommand{\theequation}{\thesubsection.\arabic{equation}}

\title{
\hfill\parbox{4cm}{\normalsize IMSC/2004/04/19\\
                               hep-th/0405119}\\        
\vspace{2cm}
Loop Variables and Gauge Invariance in Closed Bosonic String Theory.
\author{B. Sathiapalan\\ {\em Institute of Mathematical Sciences}\\
{\em Taramani}\\{\em Chennai, India 600113}\\ bala@imsc.res.in}}           
\maketitle     

\begin{abstract} 
We extend an earlier proposal for a gauge invariant description of
off-shell open strings (at tree level), using loop variables, 
to off-shell closed strings (at tree level). The basic idea is to 
describe the closed string amplitudes as a product of two open string 
amplitudes (using the technique of Kawai, Lewellen and Tye). The 
loop variable techniques that were used earlier for open strings 
can be applied here  {\it mutatis mutandis}.  It is a proposal for a 
theory whose on-shell amplitudes coincide with those of the closed 
bosonic string 
in 26 dimensions. It is also gauge invariant off-shell. As was the 
case with the open string, the interacting closed string looks like
a free closed string thickened to a band.

\end{abstract}

\newpage

\section{Introduction}

The loop variable approach described in \cite{BSLV,BSREV,BSCP} is a proposal
for a gauge invariant generalization of the world sheet sigma model 
renormalization group approach [\cite{L}-\cite{T}] to obtaining the 
tree level equations of motion of 
space-time fields in string theory. In the loop variable
 approach the fields 
are allowed to be off-shell and thus contain information about off-shell
vertices that should be used in an effective action description. 
In this sense at tree level it is an alternative to string field theory.

So far the loop variable approach has focussed on open strings.
For closed strings the free theory was outlined in \cite{BSSI} where
it was applied to the massless sector. The vertex operators for
the closed string are simple products of open string vertex operators
for the left movers and right movers with the constraint that
their dimensions have to be equal. Thus it was possible to repeat
the loop variable construction for the free open string, for the 
two sectors separately. Apart from the level matching constraint
the only communication between the two sectors was through the Liouville
mode, which need not factorize. This introduced extra terms into the equation
of motion, and these were necessary for gauge invariance. In \cite{BSSI}
this was shown for the massless fields and in this paper we will show that
these features continue to hold for all modes.
                                              
When we attempt to introduce interactions we run into a number of problems:

1.  While the vertex operators factorize, the correlation functions are
integrated over the world sheet $\int ~dz~d\zb~ <...>$ and this does
 not factorize into $\int dz$ and $\int d\zb$. Nevertheless,
it has been shown by Kawai, Lewellen and Tye (KLT) \cite{KLT} that one can continue
analytically to Minkowski signature on the world sheet 
($z \rightarrow \xi , \zb \rightarrow \eta $) and then
one can actually write the integrals as sums of products $(\int d \xi <...>)
\times (\int d\eta <...>)$ of open string amplitudes, the only coupling between
the two sectors being a phase factor. This means that one can
try to use the loop variable construction for the left movers and right movers
separately. This will be done in this paper.
 
2. The Green function $G(z, \zb ;0) ~=~ ln ~ z \zb$ splits up
conveniently into a holomorphic and anti-holomorphic part.  This is what
allows the KLT construction. However we can only use this Green function
for on-shell amplitudes. In order to go off-shell one has to use
a cut off Green function. The obvious regularization $ln ~(|z|^2+a^2)$
spoils the factorization. We will get around this by first analytically
continuing to a Minkowski world sheet and then regularizing the Green function.
Thus we get $ln ~(\xi ^2 + a^2 ) ~+~ln ~(\eta ^2 + a^2)$.
 Since $\xi ,\eta$are real variables (unlike $z,\zb$) these Green functions are 
regular for all finite values of the arguments. There is nothing wrong
with performing the renormaliztion group in Minkowski space. But it would mean
the time component of momentum has to be also regulated, it is not enough 
to regulate the invariant momenta. That is why the $\xi$ and $\eta$ parts
are separately regulated.

3. For closed strings there is also an infrared divergence because the 
range of integration is infinite. These divergences also give rise to poles
in one of the channels. We will assume that there is an infrared regulator
$R$. In the continuum infinite volume limit we expect divergences $ln ~a$ 
or $ln ~R$ to be due to poles that need to be subtracted 
out. (If we set $R~=~{l^2 \over a}$ we can  
treat both divergences in the same way.) For the gauge invariant treatment we
will keep $a,R$ finite. Gauge invariance is not affected by this. 
Both, the equations and gauge transformation laws will depend on these
parameters. This is to be expected and is the renormalization scheme
dependence that is reflected in the freedom to perform field redefinitions
of space-time fields to get different off-shell theories.

4. The fact that the Green function factorizes has another implication.
In writing down vertex operators we assume $ \p \pb X ~=~0$. (In gauge 
invariant
notation this is $\dsnmb Y ~=~0$).  This would
not be consistent if the Green function did not factorize. Thus
$<\p \pb X(z,\zb ) X(0,0)> = \p \pb ~ln~(z \zb + a^2) ~\neq ~0$. 
There are violation proportional to $a$. This creates problems when we
apply the loop variable technique because that involves Taylor expanding
the vertex operators in powers of $z$. Thus for instance the vertex operator
$\p X \pb X(z,\zb )$ would expand to give terms 
$\p X (0) \pb X (0) + \zb \p \pb X (0) \pb X(0) +...$.  This would introduce 
vertex operators that were not there in the original set. If we use the regulated
Minkowski space Green function we can always maintain the factorization property 
$\p _\xi \p _\eta G =0$. (In gauge invariant notation $\dsnmb G ~=~0$).

 As mentioned earlier the Liouville mode cannot in general be factorized.
We  expect $\p \pb \sigma (z,\zb ) \neq~0$.  This property is crucial
for gauge invariance of the equations of motions. This has to be retained 
for the gauge invariant theory where this translates to $\dsnmb \SI ~\neq~0$.

Once the abovementioned four points are taken care of, the application of 
the formalism of \cite{BSREV} to closed strings is quite straightforward:
The free closed string theory has been described in \cite{BSSI}.  All we have
to do is to replace $k_n$ by $\int ~dt~k_n(t)$ (and same for $\kb$) and $\SI$ by 
$(G~+~\SI )$. 

The gauge transformation parameter $\la _p$ gets replaced
by $\int~dt~\la _p (t)$ (and same for $\bar {\la}$). The equations 
of motion are obtained by varying w.r.t. $\SI$. Gauge invariance 
at the loop variable level is guaranteed provided we 
impose the tracelessness constraint on the gauge parameters.
 The space time gauge 
transformation  law for fields can be defined recursively as in 
the case of open strings and this ensures that the space-time field
equations are also gauge invariant.

We can argue that for on-shell physical states, in the critical dimension, 
this method gives the string theory answer. This is because of the following:

1. We know that if we set $\si$ to be a constant and choose on-shell physical 
external states, the calculation reduces to a Virasoro-Shapiro amplitude
calculation.

2. It can be shown that to linear order in 
$\si$, the replacement
$G \rightarrow G +\SI$ gives the $\si$ dependence of the partition function that one would expect
on performing a Weyl transformation on a theory that had a constant $\si$ 
to begin with:

a)  Terms in the contraction $<X(z) X(z)>$ and derivatives:
The terms of the form $\p _z ^n\sigma$ and $\p _{\bar z}^n \sigma$
can be obtained by a conformal transformation. In fact that is how they were obtained
in the loop variable approach. By assuming that $\sigma$ is a function of $z, \bar z$
we can also get the mixed derivative terms: 
$<\p X(z) \bar \p X(\bar z)> \approx \p \bar \p \sigma (z,\bar z)$.
 One can check that what is thus obtained
is in fact correct by comparing with expressions in \cite{DP} for instance. Note that
we are only interested in terms linear in $\sigma$.
 
b)$\sigma $ dependence of vertex operators: $\kn \p _z ^n X$ will get modified
to $\kn (\p _z ^n X + \p _z^{n-m} X \p _z ^m \sigma )$. In the loop variable formalism
the extra (D+1 th) coordinate provides terms of the form $q_m q_0 <\p _z ^m Y Y> \approx
q_m q_0 \p _z^m \sigma$. The combination $k_{n-m} q_m q_0$ can be set equal
to $k_n$ by some choice of gauge. Thus one expects (though this needs to be checked
in detail)
that the $\sigma$ dependence of vertex operators amounts to field redefinitions
and choice of gauge. 

In any case the issue discussed in b) does not affect the tachyon or graviton scattering 
amplitudes performed in arbitrary backgrounds (including massive modes), because these vertex
operators are not modified. Thus one should, by factorization, expect to recover
any amplitude.

Since
Weyl transformation is required to be a symmetry of the (on-shell) theory
we can get equations of motion by requiring the variation wrt $\sigma$ vanish.
All we need is the
correct linear dependence in $\sigma$, which by the above arguments, we have.

These arguments, while not rigorous, can hopefully be made rigorous. We leave this
as an issue for the future.

Thus (assuming the correctness of these arguments)
we are guaranteed to get the string theory
 answer for on-shell physical states. What this method provides is a
gauge invariant off-shell generalization of the equations. 

In the interest of conciseness, the loop variable approach for
open strings is not reviewed here. Detailed calculations illustrating
these ideas are also not given.
This paper is organized as follows: In Section II we describe the
free closed string. In Section III we elaborate on the 
four points discussed in this introduction. In Section IV we give an example
of a calculation. Section V contains some conclusions and further questions.

\section{Free Closed Strings}

We modify 
the open string loop variable to
\be    \label{lp}
exp(i\ko .X ~+~ i\int _{c} \alpha (s) k(s) \partial _{z} X(z+s, \zb + \tb ) ds +
\int _{c} \alb   (\bar{s} ) \bar{k}(\bar {s} )\partial _{\zb} X(z+s, \zb + \bar{s} )
d \bar{s})
\ee
We will Taylor expand the exponent in powers of $s, \bar{s} $, keeping
in mind that $\partial _{z} \bar{\partial _{\bar{z}}} X(z,\bar{z})=0$.
Thus the exponent becomes :
\be
\ko [X(z, \zb ) + \ai \p X + \at \pp X + \al _{3} \ppp X/2! +...
+ \aib \bar{\partial} X + \atb \bj X + \bar{\al _{3}} \bjj X/2! +...]
\ee
\[
+ \ki [ \p X + \ai \pp X + \at \ppp X /2! +..]
+ \kib [ \bar{\partial} X + \aib \bj X + \atb \bjj X /2! +...]
\]
\[
+ \kt [ \pp X + \ai \ppp X /2! +...]
+ \ktb [ \bj X + \aib \bjj X /2!+...
] +...
\]
We have assumed that $k_{0}=\bar k_{0}$.
Let
\be
Y = X + \ai \p X + \at \pp X + \al _{3} \ppp X /2! +...+
\aib \bar{\partial} X + \atb \bj X + \bar{\al _{3}} \bjj X /2! +...
\ee
Then defining $x_{n}, \bar{x_{n}}$ in the same way as before \cite{BSSI}
(${\p \al _n \over \p \xm} = \al _{n-m}$ and same for $\bar \xn$)
\be
\frac{\p Y}{\p x_{1}} = \p X + \ai \pp X + \at \ppp X /2! +...
\ee
\be
\frac{\p Y}{\p \xb _{1}}= \bar {\p} X + \aib \bj X + \atb \bjj X /2! +...
\ee
\be
\frac{\pp Y}{\p x_{1} \p \xb _{1}} = 0
\ee
The exponent becomes
\be 
\ko Y +  \sum _{n} ( k_{n}
\frac{\p Y}{\p x_{n}} +
\bar{k_{n}}
\frac{\p Y}{\p \xb _{n}})
\ee
As before we can define $<YY> =\Sigma  $  to get
\be
<\frac{\p Y}{\p x_{n}} Y> = 1/2 \frac{\p \SI}{\p x_{n}} \, ;
 <\frac{\p Y}{\p x_{n}}  \frac{\p Y}{\p x_{m}}> = 1/2
 ( \frac{\pp \SI}{\p x_{n} \p x_{m}} - \frac{\p \SI}{\p x_{n+m}} ) \, ;
\ee
\[
 <\frac{\p Y}{\p x_{n}}  \frac{\p Y}{\p \xb _{m}}> = 1/2
 ( \frac{\pp \SI}{\p x_{n} \p \xb _{m}} )
\]
The loop variable (\ref{lp}) along with its '$\Sigma $' dependence is
\be   \label{LP}
exp(i(\ko Y +  \sum _{n} ( k_{n}
\frac{\p Y}{\p x_{n}} + q_n \frac{\p Y^5}{\p \xn}) ~+~
(\bar{k_{n}} \frac{\p Y}{\p \xb _{n}} +  \bar q_n \frac {\p Y^5}{\p \bar \xn}) + 
\ee
\[
( \ko ^{2} + \qo ^{2}) \SI + (\ki . \ki
+\qi \qi ){1\over 2}(\dsii - \dst )\SI
\]
\[
+ (\ki .\ko + \qi . \qo ) \dsi \SI ~+~
( \kt . \ko + \qt \qo ) \dst \SI
+ ( \kib . \kib + \qib \qib ) {1\over 2} ( \dsiib - \dstb )\SI
\]
\[
+ ( \kib . \ko
 + \qib \qo ) \dsib \SI + ( \ktb . \ko + \qtb \qo ) \dstb \SI
+( \ki . \kib + \qi \qib ) \dsb \SI )~+~...)
\]
We have denoted the D+1 st component of '$k$' by $q$..

The equations of motion are obtained by collecting  terms of a particular
 dimension and setting the variation $\frac{\delta}
{\delta \Sigma}$ to zero.

The gauge transformations are $k(s) ~\rightarrow ~\la (s) k(s),~\bar k(\bar s )
~\rightarrow ~\bar \lambda ( \bar s ) \bar k (\bar s ) $. 
(The $q$'s also are included as the $D+1$ component of $k_n^\mu$.)
Under this transformation
one can check that the loop variable (\ref{LP}) transforms into a total
derivative provided we impose the usual tracelessness condition, familiar
from open strings : 
\[ 
<~\la _p \kn . \km ~> ~=~0, <~\la _p \kn . \bar \km ~ >~=~0
\]
\be
<~\bar {\la _p} \bar \kn . \bar \km ~> ~=~0,
 <~ \bar {\la _p} \kn . \bar \km ~ >~=~0
\ee

\subsection{Massless Fields}
\setcounter{equation}{00}

After dimensional reduction,
 the vertex operators at the lowest mass level are $\yim
\bar \yin $ and $\yim \bar Y^5_1$ and $Y^5_1 \bar Y^5_1$.
As in the case of the open strings we will impose a constraint
that removes one set of momenta. We will require that the
number of $q_1$'s is equal to the number of $\bar q_1$'s. 
This is the analog of the ghost number constraint in string field theory.
(In the open string we set $q_1~=0$ by itself, and in products,
imposed constraints to get rid of $q_1$, such as $q_1 \kim = \ktm q_0$).
The space-time fields are the graviton, the antisymmetric tensor and 
the dilaton (see (\ref{FLD}).
Thus we will ignore the equation for $\yim \bar Y^5_1$.
There are three types of terms that need to be varied for the 
massless field equations. We will focus on the equation for the
graviton - which corresponds to the vertex operator 
$\yim \bar{\yin} \e$:
\be
I. \, \,
e^{(\ko ^{2} + \qo ^{2}) \SI} \e \{ (i)^{2} \kim \kinb
\frac{\p Y^{\mu}}{\p x_{1}}
\frac{\p Y^{\nu}}{\p \xb _{1}}   \}
\ee

\be
II. \, \,
e^{(\ko ^{2} + \qo ^{2}) \SI} \e \{
(\ki . \ko + \qi \qo ) (\dsi \Sigma ) i \kimb \frac{\p Y^{\nu}}{\p \xb _{1}}
\ee
\[
+(\kib .\ko + \qib \qo ) (\dsib \Sigma ) i \kim
\frac{\p Y^{\nu}}{\p x _{1}} \}
\]

\be
III. \, \,
e^{(\ko ^{2} + \qo ^{2}) \SI } \e \{
2(\ki . \kib + \qi \qib ) 1/2 \dsb \Sigma \}
\ee

Note that for the massless fields $\qo =0$.
\be
\dds I = -(\ko^{2} + \qo^{2} ) \kim \kinb
\frac{\p Y^{\mu}}{\p x _{1}}
\frac{\p Y^{\nu}}{\p \xb  _{1}}
\e
\ee

\be
\dds II = (\ki . \ko + \qi \qo ) \kimb \kon
\frac{\p Y^{\mu}}{\p \xb  _{1}}
\frac{\p Y^{\nu}}{\p x _{1}}
\ee
\[
+(\kib . \ko + \qib \qo ) \kim \kon
\frac{\p Y^{\nu}}{\p \xb  _{1}}
\frac{\p Y^{\mu}}{\p x _{1}}
\]

\be
\dds III =
(i)^{2} \kom \kon
\frac{\p Y^{\nu}}{\p \xb  _{1}}
\frac{\p Y^{\mu}}{\p x _{1}}
(\ki .\kib + \qi \qib )
\ee
Coefficient of
\[
\frac{\p Y^{\nu}}{\p \xb  _{1}}
\frac{\p Y^{\mu}}{\p x _{1}}
\]
\be  \label{eom}
[-\ko^{2} \kim \kinb + \ki .\ko \kinb \kom + \kib . \ko \kim \kon
- \kon \kom (\ki . \kib + \qi \qib ) ] =0
\ee

We have set the $(mass)^{2}=q_{0}^{2}=0$.
The fields are the graviton, the anti symmetric tensor and the dilaton:
\[
\int [dk d\bar{k} dqd\bar{q} ] k_{1}^{(\mu } \kib ^{\nu )} \Phi [k,
\bar{k} , q, \bar{q}]
= h^{\mu \nu}
\]
\[
\int [dk d\bar{k} dqd\bar{q} ] k_{1}^{[\mu } \kib ^{\nu ]} \Phi [k,
\bar{k} , q, \bar{q}]
= A^{\mu \nu}
\]

and
\be  \label{FLD}
\int [dk d\bar{k} dqd\bar{q} ] \qi \qib \Phi [k,
\bar{k} , q, \bar{q}] = \eta
\ee

The symmetric part of (\ref{eom}) gives the linearized equation for the graviton 
and the anti-symmetric part gives the equation for the antisymmetric tensor $A^{\mu \nu}$.

The gauge transformations are
\be  \label{gt}
\ki \rightarrow \ki + \li \ko \, ; \, \kib \rightarrow \kib + \bar{\li}
\ko \, ; \, \qi \rightarrow \qi \, ; \, \qib \rightarrow \qib
\ee
Since $q_{0}$ is zero, $q_{1}, \bar{q_{1}}$ remain invariant.  This
gives the usual infinitesimal transformation for the graviton
\be
h^{\mu \nu}   \rightarrow
h^{\mu \nu}    +
 \p ^{(\mu} \epsilon ^{\nu )}  + \p ^{(\nu} \bar{\epsilon }^{\mu )} =
h^{\mu \nu}    + \p ^{(\mu} \epsilon _{S}^{\nu )}
\ee
where we define
\be
\epsilon _{S}^{\mu} = \epsilon ^{\mu} + \bar{\epsilon} ^{\mu}
\ee
and
\be
A^{\mu \nu} \rightarrow A^{\mu \nu } + \p ^{[ \mu } \epsilon ^{\nu ]}
+ \p ^{[ \nu } \bar{\epsilon}^{\mu ]} = A^{\mu \nu} + \p ^{[\mu}
\epsilon _{A} ^{\nu]}
\ee
where \be
\epsilon _{A}^{\mu} = \epsilon ^{\mu} - \bar{\epsilon} ^{\mu}
\ee
and the dilaton
\be
\delta \eta = 0
\ee
One can easily check that under the variations (\ref{gt}) equation
(\ref{eom}) is invariant.  

\subsection{Massive Fields}
\setcounter{equation}{00}

From the point of view of the D+1 - dimensional theory all fields are massless.
So there is nothing new in this calculation. We just repeat what was
done for level (1,1) above, for the level (2,2) operators. These
are $\yim \yin \bar Y_1^\rho \bar Y_1 ^\sigma$, $\yim \yin \bar Y_2^\rho$,
$\ytm \bar \ytn$.  

\[
e^{i\ko .Y ~+~\ko ^2 \SI }\{ [i\kt Y_2 ~-~ {1\over 2!}\ki . Y_1 ~\ki . Y_1
+ \kt .\ki \dst \SI + \ki .\ki {1\over 2} (\dsii ~-~ \dst )\SI +
\]
\[
 i\ki Y_1 ~\ki .\ko \dsi \SI ][ ~anti-holomorphic ~] +
i\kt . \ktb {\pp \over \p x_2 \p \bar x_2 }\SI ~+
\]
\be   \label{lev2}
- \ki .Y_1 ~\kib .\bar Y_1 ~\ki .\kib  {\pp \over \p x_1 \p \bar x_1}\SI ~+~
~i\kib .\bar Y_1 ~\kt . \kib {\pp \over \p x_2 \p \bar x_1 }\SI~+
~i\ki .Y_1 ~\ki . \ktb {\pp \over \p x_1 \p \bar x_2 }\SI ~\}
\ee

Let us evaluate the coefficient of $\ytm \bar \ytn$ in the equation of motion.
Varying w.r.t. $\SI$ gives:

\[
[- \ko ^2 \ktm \bar \ktn ~+~ \ktm \ktb . \ko \kon ~-~ \ktm \kib .\kib \kon
~+~ \ktm \bar \kin \kib . \ko ~+~\bar \ktn \kt .\ko \kom 
\]
\be      \label{eom2}
~-~\bar \ktn \ki .\ki \kom ~+~ \bar \ktn \kim \ki .\ko 
~-~ \kt .\ktb \kom \kon ~-~ \ki .\kib \kim \bar \kin ~-~
 \kt .\kib \kom \bar \kin ~-~ \ki .\ktb \kim \kon ]
\ytm \bar \ytn ~=~0
\ee

The gauge transformations

\be    \label{gt2}
\ktm  ~\rightarrow~\ktm ~+~ \la _1 \kim ~+~ \la _2 \kom 
\ee

have to be added to (\ref{gt}) 
(with the $q$ as the $D+1$th component of $k_n^\mu$).
 One can check that the
equation of motion (\ref{eom2}) is invariant, provided one uses
the tracelessness condition on the gauge parameters.

After dimensional reduction $\qo ^2 =2$. We do not give the
details here.

\section{Interactions}
\setcounter{equation}{00}

\subsection{Analytic Continuation and KLT prescription}

We will assume that the world sheet has been analytically
continued: $z~=~ x ~+~iy ~=~ \xi$ with $y=y_1+iy_2$. In the complex $y$ plane
we rotate the integration contour from the $y_1$-axis to the $y_2$-axis
 as explained
in \cite{KLT,BSFRG}\footnote{The application of the KLT prescription to the
renormalization group calculation in closed strings is described in \cite{BSFRG}.}.
 Thus $\xi ~=~ x~-~y_2$ on the rotated contour. Similarly $\zb ~\rightarrow ~\eta
~= x~+~y_2$. $\xi ,\eta$ are real on the rotated contour and 
$\int ~ d^2z ~=~\int ~ d\xi ~d\eta $ factorizes, as do the correlation 
functions.  The range of integration can formally be taken to be $-R$ to $+R$.
But while evaluating the integrals the phase factors mentioned below have to be 
incorporated. 
We use $ln ~(\xi ^2 ~+~ a^2)~$ and $ln ~(\eta ^2 + a^2)$ 
as the regulated Green functions. 

Interactions involve several vertex operators. By Taylor expanding all the vertex operators 
about one point, the loop variable expression formally looks like that of the free theory
with the loop now thickened to a band. This can then be easily 
made gauge invariant \cite{BSGI,BSREV}.
As explained in the above papers the continuum limit can be taken only after resumming
the Taylor expansion series. Otherwise, since each term has inverse powers of the short-distance
cutoff $a$, the limit $a\rightarrow 0$ cannot be taken.

We expand the Green function in powers of ${\xi \over a} , {\eta \over a}$.
This is equivalent to Taylor expanding the vertex operators about the origin:
\be
X(\xi ,\eta)  = X(0,0) + \sum _{n=1} ({\xi ^n \over n!}\p ^n X(0) + {\eta ^n \over n!} \bar \p ^n X(0))
\ee
(Note that we have used the fact that $\p \bar \p X =0$.) 

Thus 
\[
\int d\xi \int d\eta \ko (\xi ,\eta ).X(\xi ,\eta ) = \int d\xi \int d\eta \ko (\xi, \eta ) .X(0,0) ~+
\]
\[ \sum _{n=1} \{
\int d\xi \ko (\xi ). {\xi ^n\over n} \yn (0) + \int d\eta \ko (\eta ).{\eta ^n \over n} \bar \yn (0)\}
\] 
where $\yn = {\p ^n X\over (n-1)!} = {\p Y\over \p \xn}|_{\xn=0}$ and $\ko (\xi ) \equiv \int d\eta \ko (\xi ,\eta )$
and $\ko (\eta ) \equiv \int d\xi \ko (\xi ,\eta )$.

As in the open string, we can write 
\[
\ko .Y (\xi ,\eta ) + \sum _{n=1} [ \kn (\xi ) \yn (\xi ) + \bar \kn (\eta ) \bar \yn (\eta )]
= \ko . Y (0) + \sum _{n=1}[ K_n (\xi ) \yn (0) + \bar K_n (\eta ) \bar \yn (0) ]
\]
and 
all $\xi$ dependences can be lumped into
\be   \label{int}
K_q (\xi ) = \sum _{n=0}^{n=q} k_n (\xi)  D_{n}^{q}(\xi )^{q-n}
\ee
where
\br
D_n^q &  = & ^{q-1}C_{n-1},\; \; n,q\ge 1 \nonumber \\
      &  = & {1\over q}, \; \; n=0 \nonumber \\
      &  = & 1 ,   \;\;           n=q=0
\er
and similarly for $\eta$.
Note that in $\kn$,  $\xi$ is a label denoting the location, whereas
in $K_n$ there is an explicit functional dependence as given in (\ref{int}).
The point of doing this is that expressed in terms of $ \int ~d \xi ~K_n (\xi )$,
the interacting theory looks exactly like a free theory. 
 Once this is done the technique of obtaining gauge invariant 
equations for the free theory, described in the last section can be directly
applied here.

As mentioned in the introduction we will use (\ref{LP}) with $\si$ replaced
by $G~+~\si $. Here we will take $G (\xi, \eta ;0,0) ~=~ ln ~(\xi ^2 + a^2) ~+~
ln ~ (\eta ^2 +a^2)$.  Note that ${{\pp G}\over {\p \xi \p \eta}}~=~0$
whereas ${{\pp \si}\over {\p \xi \p \eta}}~\neq ~0$, also as mentioned in the
introduction. After introducing $\xn , \bar \xn$, this will become $\dsnmb G ~=~0$
and $\dsnmb \SI ~\neq ~0$. 

There is one further term in the Green function that is very important.
We must add to $k_{0i}.k_{0j} G(\xi _i, \eta _i ; \xi _j, \eta _j )$ the term
$i\pi k_{0i}.k_{0j}\theta (-(\xi _i - \xi _j )(\eta _i - \eta _j))$. 
This gives rise
to the very important relative phases of the contributions from the
various integration regions. This comes from the choice of Riemann sheet
when one goes around the  branch point \cite{KLT}. 
Since this is multiplied by $k_{0i} .k_{0j}$ only and does not contribute in
terms of the type $k_n .k_m \dsnm G $, we can just add it to $G$ without
dressing it up with $\xn$'s.  The property of the loop variable that
it changes by a total derivative in $\xn$  on performing a gauge transformation 
with $\la _n$ is not affected by an overall $\xn$-independent multiplication
factor. However this extra factor will have to be included when one constructs
the map from loop variables to space-time fields which is then used to define the
gauge transformation properties of the fields. Similarly when the generalized
tracelessness constraint that is imposed on loop variables is mapped to a constraint
on space time fields, this factor must be included. $ G(\xi _i, \eta _i ; \xi _j, \eta _j)$
is Taylor expanded in powers of $\xi ,\eta $ but $\theta $-function will of course be
left intact. In practice this will imply that the integration regions of $\xi _i , \eta _j$ will have
to be broken up and phase factors attached appropriately to each region.

\subsection{Symmetries}

Gauge transformations, in the interacting case, will be given by

\be   \label{GTI}
\int _{-R}^R ~d\xi K_n (\xi ) 
~\rightarrow ~\int _{-R}^R ~d\xi K_n (\xi ) 
~+~ \sum _{p=1}^{n-1}~\int _ {-R}^R~d\xi ' ~\la _p (\xi ') \int _{-R}^R~d\xi ~K_{n-p} (\xi )
\ee

Of course there is a correponding gauge transformation for the ``bar'' variables.

Note that in addition to this, there is a ``global'' Lorentz invariance on the vector index. 
$\kn ^\mu ~\rightarrow~ \Lambda ^{\mu \nu } \kn ^\nu$. Although in terms of loop variables
it looks like a global rotation,  $\Lambda ^{\mu \nu}$ is any
antisymmetric tensor, and can be a function of  $\ko$, $\la _n$ or even $\kn$. Thus
in  space time, it is a local Lorentz transformation. This is only possible because of 
the presence of massive fields that provide ``connection''-like terms that ensure invariance.
This can be seen when one defines the Lorentz transformation on the space time fields.

\subsection{Mapping to Space-Time Fields}

Until one maps to space-time fields the loop variable method gives a formal expression
- a ``sentence'' of ``words'' composed out of an alphabet made up of $\kn $ and $\bar \kn$ and the 
coefficient of each ``word'' is given by an integral (with phase factors taken into account).
After performing the integrals one gets some functions of $R/a$ as the coefficients. If one makes
the replacement given in (\ref{GTI}), one finds the ``sentence'' is unchanged. After mapping 
to space-time fields this acquires an interpretaion as the gauge invariance of an equation of motion.
The crucial question is whether one can extract well defined gauge transformation for space-time fields
from the abstract expression given in (\ref{GTI}). This question was answered in the affirmative
in \cite{BSGI} for open strings. A recursive algorithm was given for assigning gauge transformation
properties to space time fields. A ``word'' (W) is mapped to an expression involving space-time fields  (F) 
and its gauge
transformed ``word'' (W') is also mapped to an expression (F') involving space-time fields. 
The transformation of space time fields should thus change F to F'. That such transformations can be defined
was shown recursively. 
It is recursive in the level of the field, i.e., transformation of fields
at higher level were fixed by (\ref{GTI}) in terms of fields at lower levels.  The ``words'' (W) that are
mapped to space-time fields in this method are precisely the ones that are there in the ``sentences'' (equations of motion).
Thus the invariance of the equation of motion (expressed in terms of space-time fields) follows
{\it a fortiori} from the invariance of the ``sentences''. 

One can 
apply the same technique for the closed strings also. The level matching constraint 
($L_0 = \bar L_0$) ensures that one can order by level here also. 

There is one crucial difference.
The phase factor must be included in our choice of ``words'' (W) 
that  we map  to expressions (F) involving space-time fields in order to extract the
gauge transformation properties of fields.  A simple way to see that this is to imagine extracting 
the gauge transformation properties
of space-time fields with and without the phase factor. Both are well defined procedures,
 but they give different answers. It is easy to see that the equations of motion, which
do have the phase factors in them, will be invariant under the first one but there is no reason
for them to be invariant under the second one. 

The map itself is explained best by examples:
\[   
 \int d\xi _i \int d \eta _j
<k_{ni}^\mu (\xi _i) \bar k_{mj}^\nu (\eta _j)>  = 
 \int d\xi _i \int d \eta _j
S_{n;n}^{\mu \nu }(\xi _i , \eta _j ) \delta _{n,m}
\]
\[
\int d\xi _1 \int d \xi _2 \int d \eta _3 \int d \eta _4
<k_{n1}^\mu (\xi _1) k_{n2}^\nu (\xi _2) \bar k_{m3}^\rho (\eta _3)\bar k_{m4}^\sigma (\eta _4)>  = 
\]
\[
\int d\xi _1 \int d \xi _2 \int d \eta _3 \int d \eta _4 \{
S_{n_1,n_2;m_3,m_4} ^{\mu \nu \rho \sigma} (\xi _1 ,\eta _3) 
\delta (\xi _1 -\xi _2) \delta (\eta _3 -\eta _4) \delta _{n_1+n_2, m_3 +m_4} ~+
\]
\[
S_{n_1;m_3}^{\mu \rho }(\xi _1 , \eta _3 )\delta _{n_1,m_3}  
S_{n_2;m_4}^{\mu \sigma }(\xi _2 , \eta _4 )\delta _{n_2,m_4} ~+
\]
\be  \label{Map1}
S_{n_1;m_4}^{\mu \sigma }(\xi _1 , \eta _4 )\delta _{n_1,m_4}  
S_{n_2;m_3}^{\mu \rho }(\xi _2 , \eta _3 )\delta _{n_2,m_3}\} 
\ee
Note that because of the factorization property, {\it a priori} 
there is nothing that connects a particular $\xi _i$ with
a $\eta _j$. That $\xi _i ,\eta _j$ represent the coordinates of
one vertex operator becomes evident once the two get ensconced inside a field
$S(\xi _i , \eta _j)$. This is effectively a property of the zero mode
$\ko (\xi _i , \eta _j)$, which through a relation of the form 
\[
S(X) = \int d\ko (\xi _i,\eta _j) S(\ko (\xi _i ,\eta _j))e^{i\ko (\xi _i ,\eta _j) X}
\]
 relates the field with the vertex 
operator location. 
The assignment
of $\xi _i$ with some $\eta _j$ is also  implicit in the phase factor which involves terms
 of the form
$\ko (\xi ,\eta ).\ko (\xi ',\eta ')\theta (-( \xi - \xi ')(\eta -\eta '))$
where it is understood that $\xi ,\eta$ refer to one vertex operator and $\xi ',\eta '$
refer to another vertex operator. The theta function will be defined to be zero when $\xi$, $\xi '$
and $\eta, \eta '$ 
refer to the same vertex operator so that both $\xi = \xi '$ and $\eta = \eta '$ .

If we include phase factors and powers of momenta in the above map we get
for example
\[
\int d\xi _1 \int d \xi _2 \int d \eta _3 \int d \eta _4 
 <e^{i\pi \int d\xi \int d\eta \int d\xi ' \int d\eta '\ko (\xi ,\eta ).\ko (\xi ' ,\eta ')\theta (-(\xi -\xi ' )(\eta - \eta '))}
\]
\[
k_{n1}^\mu (\xi _1) k_{n2}^\nu (\xi _2) \bar k_{m3}^\rho (\eta _3)\bar k_{m4}^\sigma (\eta _4)>  = 
\]
\[
\int d\xi _1 \int d \xi _2 \int d \eta _3 \int d \eta _4 \{
S_{n_1,n_2;m_3,m_4} ^{\mu \nu \rho \sigma} (\xi _1 ,\eta _3) \delta (\xi _1 -\xi _2) \delta (\eta _3 -\eta _4) \delta _{n_1+n_2, m_3 +m_4} ~+
\]
\[
e^{i\pi \ko (\xi _1,\eta _3 ).\ko (\xi _2 ,\eta _4)\theta (-(\xi _1 -\xi _2 )(\eta _3 - \eta _4))}
S_{n_1;m_3}^{\mu \rho }(\xi _1 , \eta _3 )\delta _{n_1,m_3}  
S_{n_2;m_4}^{\mu \sigma }(\xi _2 , \eta _4 )\delta _{n_2,m_4} ~+
\]
\be   \label{Map3}
e^{i\pi \ko (\xi _1 ,\eta _4).\ko (\xi _2 ,\eta _3)\theta (-(\xi _1 -\xi _2 )(\eta _4 - \eta _3))}
S_{n_1;m_4}^{\mu \sigma }(\xi _1 , \eta _4 )\delta _{n_1,m_4}  
S_{n_2;m_3}^{\mu \rho }(\xi _2 , \eta _3 )\delta _{n_2,m_3} \}
\ee
Implicit in this example is the rule
\be
<\kom (\xi ,\eta ) k_{n}^\nu (\xi _1) \bar k_m^\rho  (\eta _2)> = \kom (\xi _1 ,\eta _2 )S_{n,m}^{\nu \rho}
(\xi _1 ,\eta _2) \delta (\xi -\xi _1)\delta (\eta - \eta _2)\delta _{n,m}
\ee
Thus in example (\ref{Map3}) in one of the terms involving two fields, $\eta _3$ is
 associated with $\xi _1$ and  $\eta _4 $ with $\xi _2$, and vice versa in the other term.

The general rule is clear from the above. All possible contractions of the 
$\kn, \bar \kn$ consistent with level matching, are made.
This fixes the association of the $\eta$'s and $\xi$'s with each other.

Let us proceed to an example that illustrates some of these ideas.

\section{Example}

In \cite{BSFRG} some terms in the on-shell transverse three graviton vertex was 
calculated. We will repeat the same calculation here in the gauge invariant case.
We will see how the leading term in the Taylor expansion is reproduced.  They are
of course accompanied by terms involving massive modes. These are required for the
full gauge invariance of the theory. 

We look for a term proportional to $ Y_1^\rho  \bar \yin$. Since for the propagator
we are using, $G_{1,0}(0) =0$, the leading term involves $G_{2,0} (0),G_{1,1}(0)$:

\[
\ko (\xi _5, \eta _5 ). \ko (\xi _6,\eta _6)e^{\int \int 
\ko (\xi ,\eta ).\ko (\xi ', \eta ')[G(0) + i\pi \theta (-(\xi - \xi ')(\eta - \eta '))]}
\]

\[
[K_2(\xi _1) .\ko (\xi _2) G_{2,0} + {1\over 2} K_1(\xi _1 ). K_1(\xi _2) G_{1,1}]
[\bar K_2(\eta _1) .\ko (\eta _2) \bar G_{2,0} + {1\over 2} \bar K_1(\eta _1 ).\bar K_1(\eta _2) \bar G_{1,1}]
\]
\be  \label{3G}
 K_1^\rho (\xi _3) \bar K_1^\nu (\eta _3) Y_1^\rho (0) \bar \yin (0)
\ee

Integrals over locations $\xi _i ,\eta _j$ are understood.

\[
K_1(\xi ) = \ki (\xi ) + \xi \ko (\xi )
\]
\be
K_2 (\xi ) = \kt (\xi ) + \xi \ki (\xi ) + {\xi ^2 \over 2} \ko (\xi )
\ee

Substitue these in (\ref{3G}) and use $G_{1,1} = -G_{2,0} = {1\over a^2}$.
Keep the term of the form $\ki .\ko$ in the contractions and $k_1^\rho \bar \kin$ in the
coefficient of $Y_1^\rho \bar \yin$. Perform the contractions:
\[
<\kim (\xi _1) \bar \kin (\eta _3)> = h_{1;1}^{\mu \nu}(\xi _1 ,\eta _3)
\]
\[
<\kim (\xi _3) \bar \kin (\eta _1)> = h_{1;1}^{\mu \nu}(\xi _3 ,\eta _1)
\]
The first square brackets gives
\[
{1\over a^2}[\xi _1 \ki (\xi _1).\ko (\xi _2 ) - \ki (\xi _1 ).\ko (\xi _2 ) \xi _2] \bar \kin (\eta _3)
= {\xi _1 -\xi _2 \over a^2}[\ki (\xi _1) .\ko (\xi _2 )]\bar \kin (\eta _3 )
\]
If $\xi _2 = \xi _1$ we get a longitudinal graviton, which has no counterpart 
(being zero for physical states) in the
on-shell calculation done in \cite{BSFRG}. So we take the case
$\xi _2 = \xi _3$. We get a very similar expression from the second square bracket term.   
Finally the prefactor $\ko .\ko$ in (\ref{3G}), 
on dimensional reduction gives a piece $\qo .\qo \approx 2$
which is the renormalization group dimension of this term.
Putting all this together gives the quadratic term involving two gravitons in the equation of motion.

\be  \label{GG} 
 {(\xi _1 -\xi _3 )\over a^2} {(\eta _1 -\eta _3 )\over a^2}
[\ki (\xi _1) .\ko (\xi _3 )]\bar \kin (\eta _3 )
[\bar \ki (\eta _1) .\ko (\eta _3 )] \ki ^\rho (\xi _3 )
\ee
We see that this corresponds to keeping the leading term in the Taylor
expansion of the Green function 
\[
<\p _\xi X^\mu (\xi _1)X^\nu (\xi _2) > = -{g^{\mu \nu}(\xi _1 - \xi _2 )\over 4((\xi _1 - \xi _2 ) ^2 + a^2)}
\]
\[
=-{g^{\mu \nu}(\xi _1 -\xi _2) \over 4a^2}[1- {(\xi _1 -\xi _2)^2\over a^2} + ...]
\]
in the contraction of $X$'s in the OPE of two graviton vertex operators, calculated
in \cite{BSFRG}:
\be
\int ~d^2z~ \kim \kinb :\p _z X^\mu \p _{\bar z} X^\nu e^{i\ko .X}:
\int ~d^2w~ p_1^\rho \bar p _1^\sigma :\p _w X^\rho \p _{\bar w} X^\sigma e^{ip_0 .X}:
\ee

The term we are focusing on in the OPE is:
\be  \label{OPE}   
[\kim p_1^\rho {1\over 16|z-w|^2}p_0^\mu \kinb \bar p_1^\sigma \ko ^\sigma]
 :\p _w X^\rho \p _{\bar z} X^\nu e^{i\ko X(z) + p_0 X(w)}:|z-w|^{\ko .p_0\over 2}
\ee
 with ${1\over z-w}$ replaced by $\xi _1 - \xi _3 \over (\xi _1 - \xi _3)^2+a^2$ and 
 with ${1\over \bar z- \bar w}$ replaced by $\eta _1 - \eta _3 \over (\eta _1 - \eta _3)^2+a^2$. 
We also set 
\br
\ki ^\rho (\xi _3) &=& p_1^\rho \\ \nonumber 
\kim (\xi _1) &=& \kim \\ \nonumber 
\bar \kin (\eta _3) &=& \bar \kin  \\ \nonumber 
\bar \ki ^\sigma (\eta _1) &=& \bar p_1^\sigma \\ \nonumber 
\er
We see that (\ref{OPE}) gives the same result as (\ref{GG}).

Of course we must include the all important phase factor in (\ref{3G}) before performing
the integrals in Minkowski world sheet.  This calculation shows how the infinite
series of terms coming in the gauge invariant calculation are related to (the Taylor expansion of)
one term in the on-shell gauge fixed calculation.

This concludes the discussion of the two graviton term. We give an example of the massive mode
contribution to the equation of motion:
Consider the $K_2 \bar K_2 K_1 \bar K_1 G_{2,0} \bar G_{2,0}$ term. If we keep the leading term
we get
\[
\kt (\xi _1).\ko (\xi _2) \ktb (\eta _1)\ko (\eta _2) \kim (\xi _3) \kinb (\eta _3)
\]
This gives
\[
\{S_{2,1;2,1}^{\rho \mu \sigma \nu} \ko ^\rho  \ko ^\sigma 
\delta (\eta _1-\eta _3) \delta (\eta _1 -\eta _2)
\delta (\xi _1-\xi _3) \delta (\xi _1 -\xi _2)
+\]
\[
 \ko ^\rho \ko ^\sigma [S_{2;2}^{\rho \sigma} (\xi _1 ,\eta _1)
h_{1,1}^{\mu \nu} (\xi _3 ,\eta _3)] e^{i\pi \ko (\xi _1).\ko (\xi _3)
\theta (-(\xi _1-\xi _3)(\eta _1 -\eta _3))}\}G_{2,0}\bar G_{2,0} 
\]
as the leading term involving massive modes.
 Integrals over world-sheet coordinates from $-R$ to $+R$ are understood.

\section{Conclusions}

In this paper we have extended the proposal of \cite{BSGI,BSREV,BSCP}
to closed strings. The crucial ingredient is the 
prescription of KLT for writing closed string tree amplitudes
in terms of open string ones. Thus the loop variable trick
of introducing interactions by thickening the loop to a band
works for the closed string also. The gauge symmetries continue
to have the interpretation of space-time  scale transformations that
are not constant along the string.

There are many issues that still need to be understood. A 
detailed proof of the equivalence with string theory along the lines
indicated in the introduction would put this proposal on a firm
footing. The low energy symmetries - general coordinate invariance (Yang-Mills for open strings)
 - should be understandable in this formalism because of the close connection
with the sigma model. The physical significance of the various ingredients
that went into this construction remain to be understood. In particular,
the extra dimension, the method of introducing interactions, and the form of the
gauge transformations. The extension to loop amplitudes is an obvious open question.
It would be interesting to see if modular invariance admits a simple space-time interpretation
in this formalism.
Finally, at a practical level, the utility of this approach
for generating solutions still needs to be demonstrated.

\end{document}